\documentclass{bmvc2k}

\usepackage[utf8]{inputenc} 
\usepackage[T1]{fontenc}    
\usepackage{hyperref}       
\usepackage{url}            
\usepackage{booktabs}       
\usepackage{amsfonts}       
\usepackage{nicefrac}       
\usepackage{microtype}      
\usepackage{xcolor}         
\usepackage{graphicx}
\usepackage{multirow}
\usepackage{amsmath} 

\usepackage{pifont}

\newcommand{\cmark}{\ding{51}}
\newcommand{\xmark}{\ding{55}}

\title{SpecBPP: A Self-Supervised Learning Approach for Hyperspectral Representation and Soil Organic Carbon Estimation}

\addauthor{Daniel La'ah Ayuba}{d.ayuba@surrey.ac.uk}{1}
\addauthor{Jean-Yves Guillemaut}{j.guillemaut@surrey.ac.uk}{1}
\addauthor{Belen Marti-Cardona}{b.marti-cardona@surrey.ac.uk}{2}
\addauthor{Oscar Mendez Maldonado}{o.mendez@surrey.ac.uk}{1}

\addinstitution{
 Center for Vision, Speech, and Signal Processing\\
 University Of Surrey\\
 Guildford, UK
}
\addinstitution{
 Centre for Environmental Health and Engineering\\
 University Of Surrey\\
 Guildford, UK
}

\runninghead{Daniel et al}{SpecBPP}

\begin{document}

\maketitle

\begin{abstract}
  Self-supervised learning has revolutionized representation learning in vision and language, but remains underexplored for hyperspectral imagery (HSI), where the sequential structure of spectral bands offers unique opportunities. In this work, we propose \textit{Spectral Band Permutation Prediction (SpecBPP)}, a novel self-supervised learning framework that leverages the inherent spectral continuity in HSI. Instead of reconstructing masked bands, SpecBPP challenges a model to recover the correct order of shuffled spectral segments, encouraging global spectral understanding. We implement a curriculum-based training strategy that progressively increases permutation difficulty to manage the factorial complexity of the permutation space. Applied to Soil Organic Carbon (SOC) estimation using EnMAP satellite data, our method achieves state-of-the-art results, outperforming both masked autoencoder (MAE) and joint-embedding predictive (JEPA) baselines. Fine-tuned on limited labeled samples, our model yields an \textit{R² of 0.9456, RMSE of 1.1053\%}, and \textit{RPD of 4.19}, significantly surpassing traditional and self-supervised benchmarks. Our results demonstrate that spectral order prediction is a powerful pretext task for hyperspectral understanding, opening new avenues for scientific representation learning in remote sensing and beyond.
\end{abstract}


\section{Introduction}

Hyperspectral imagery (HSI) provides rich spectral information across hundreds of contiguous wavelength bands, offering powerful capabilities for material characterization on Earth's surface. A vital application is estimating soil organic carbon (SOC) content, a key indicator of soil health and a significant component of the global carbon cycle \cite{Lal2004}. Despite the data-rich nature of HSI, obtaining ground-truth SOC labels remains costly and labor-intensive, creating a label-scarce domain where unlabeled hyperspectral data are plentiful but labeled examples are rare \cite{Paoletti2019}. This gap motivates self-supervised learning approaches to leverage abundant unlabeled HSI for robust representation learning.

Self-supervised learning (SSL) has transformed representation learning by devising \textit{pretext tasks} that require no manual labels. Current approaches include masked image modeling, where models reconstruct masked regions \cite{he2022masked}, and contrastive/joint embedding methods that learn invariant representations \cite{chen2020simple, assran2023ijepa, rs16183399}. However, applying these techniques directly to hyperspectral data has limitations: masked reconstruction primarily captures local spectral correlations, while contrastive methods require careful augmentation design that is challenging for HSI data. Crucially, existing methods fail to explicitly leverage the sequential structure \cite{Goetz2009} of hyperspectral data, where bands follow the electromagnetic spectrum's natural ordering and exhibit strong correlations with adjacent wavelengths.

We propose \textit{Spectral Band Permutation Prediction (SpecBPP)}, a novel self-supervised approach tailored to hyperspectral data. SpecBPP treats each spectrum as a sequence, partitioning it into multiple contiguous segments, shuffling these segments, and training the model to predict the correct original order. This "spectral jigsaw puzzle" forces the model to learn the natural ordering \cite{Shaw2003} of wavelength regions and to capture long-range dependencies across the spectrum. Unlike masked autoencoding that primarily exploits local smoothness, our permutation prediction task encourages global understanding of spectral signatures, compelling networks to identify characteristic absorption features and place them in the correct spectral context. 

To address the factorial growth of possible permutations, we implement a curriculum learning strategy \cite{Bengio2009, Hacohen2019, Soviany2022}, that begins with simpler permutation tasks (fewer segments) and progressively increases complexity as the model learns. This staged approach helps the network gradually master spectral ordering from coarse to fine-grained patterns, making the learning process more efficient and stable.

Empirically, SpecBPP yields state-of-the-art results for SOC prediction. By pretraining a hyperspectral encoder on unlabeled HSI data using our self-supervised task and fine-tuning on limited SOC-labeled samples, we achieve representations that significantly outperform conventional SSL methods and supervised baselines, especially in low-label situations. Our contributions include:

\begin{itemize}
\item \textbf{Novel spectral permutation SSL for HSI:} We introduce the first self-supervised framework exploiting spectral ordering as a supervisory signal for hyperspectral representation learning.
\item \textbf{State-of-the-art SOC estimation:} Using SpecBPP pretraining, we improve prediction accuracy over existing approaches, achieving R² of 0.9456, RMSE of 1.1053\%, and RPD of 4.19.
\item \textbf{Improved generalization and interpretability:} The learned representations transfer robustly to scenarios with limited labeled data while providing interpretable features that correlate specific spectral regions with SOC content.
\item \textbf{Bridging ML and environmental science:} Our approach incorporates domain knowledge about spectral ordering and key wavelength features to advance SOC mapping, supporting carbon monitoring and climate-smart soil management.
\end{itemize}

\section{Related Work}
\label{related_work}

\textbf{Self-Supervised Visual Representation Learning:}
Self-supervised learning has emerged as a powerful paradigm for learning visual features from unlabeled data. Contrastive approaches like SimCLR \cite{chen2020simple} and MoCo \cite{He2020} learn representations by maximizing agreement between augmented views of the same image while distancing different images. Negative-free methods such as BYOL \cite{Grill2020} and SimSiam \cite{ChenHe2021} mitigate representational collapse using asymmetric networks and stop-gradient operations. Another successful approach is masked prediction: Masked Autoencoders (MAE) \cite{he2022masked} reconstruct randomly masked image patches, forcing models to capture contextual information. More recently, I-JEPA \cite{assran2023ijepa} operates in feature space by predicting representations of masked regions, avoiding pixel-level reconstruction while maintaining competitive performance. These advances provide foundational techniques that we adapt for hyperspectral data.

\noindent \textbf{Ordering and Permutation-Based Pretext Tasks:}
Several self-supervised methods leverage intrinsic data ordering as a supervisory signal. In computer vision, solving jigsaw puzzles (rearranging permuted image patches) \cite{Noroozi2016} and predicting relative positions of image regions \cite{Doersch2015} enable learning of spatial context. For sequential data like video, models learn temporal structure by verifying frame order \cite{Misra2016} or detecting shuffled sequences \cite{Fernando2017}. Similarly, NLP models like BERT \cite{Devlin2019} and XLNet \cite{Yang2019} incorporate order-based pretraining tasks such as next-sentence prediction and permutation language modeling. These methods share a common principle: shuffling natural data ordering (spatial patches, temporal frames, or text segments) to create self-supervised tasks that require understanding of high-level structure. Our work extends this paradigm to the spectral dimension of hyperspectral imagery, leveraging the natural ordering \cite{Shaw2003} of the electromagnetic spectrum.

\noindent \textbf{Self-Supervised Learning for Hyperspectral Imagery:}
Recent work has begun adapting self-supervised techniques to hyperspectral image (HSI) analysis, where labeled data is often scarce. Contrastive learning approaches \cite{hou2021hyperspectral, hu2021contrastive, guan2022cross, zhang2022cross} for HSI define positive pairs through spectral and spatial augmentations, while masked modeling methods \cite{liu2023self, kong2023instructional, li2024demae, wang2024hsimae} reconstruct masked portions of the hyperspectral cube. Hybrid approaches combine contrastive and reconstruction objectives to leverage the unique characteristics of HSI data \cite{huang2023spectral, cao2023transformer, cao2024mask}. These methods explore various pretext tasks from spectral clustering to band prediction, aimed at learning transferable spectral-spatial features. Our approach contributes to this emerging field by introducing spectral permutation prediction, which explicitly leverages the sequential structure \cite{Goetz2009} of spectral bands in a way previous HSI self-supervised methods have not explored.

\noindent \textbf{Soil Organic Carbon Estimation with Hyperspectral Data:}
Soil Organic Carbon (SOC) estimation using hyperspectral data has traditionally relied on supervised calibration models trained on field-collected samples with laboratory-measured SOC values. Studies have demonstrated the feasibility of mapping SOC from proximal, airborne, and satellite hyperspectral imagery \cite{Stenberg2010, Stevens2013}. Traditional approaches use methods like partial least squares regression (PLSR) \cite{gholizadeh2017leaf}, while recent work explores spaceborne hyperspectral sensors like EnMAP \cite{Bouslihim2024} and PRISMA \cite{reddy2024integrating} for regional SOC mapping. Advanced techniques including deep learning have shown promise \cite{datta2022soil}, but supervised approaches remain limited by the need for extensive soil sampling and struggle with generalization across diverse soil conditions. This creates motivation for self-supervised methods that can leverage unlabeled hyperspectral data to learn generalizable spectral features. Our SpecBPP approach addresses this need by learning soil-relevant representations through a novel spectral permutation task, bridging hyperspectral self-supervised learning and practical soil property estimation.

\section{Spectral Band Permutation Prediction (SpecBPP)}
We introduce \textbf{Spectral Band Permutation Prediction (SpecBPP)}, a novel self-supervised framework that leverages the sequential structure of electromagnetic spectra \cite{Goetz2009, Shaw2003}. This section formalizes our task definition, details the model architecture, specifies training objectives, and explains our curriculum learning strategy for managing permutation complexity.

\subsection{Task Definition}

We formalize \textbf{Spectral Band Permutation Prediction} as a self-supervised ordering task. Let $x \in \mathbb{R}^{B}$ be a spectral signature with $B$ bands (EnMAP: $B=224$). We partition $x$ into $N$ disjoint segments $x^{(1)}, x^{(2)}, \dots, x^{(N)}$, each spanning $\ell = B/N$ bands. A random permutation $\pi \in S_N$ reorders these segments to produce a permuted signature:
\begin{equation}
\hat{x} = \pi(x) = \big[x^{(\pi(1))},\; x^{(\pi(2))},\; \dots,\; x^{(\pi(N))}\big]
\end{equation}
where $\pi(i)$ gives the original segment index at position $i$ in $\hat{x}$. Since $\pi$ is a bijection, its inverse $\pi^{-1}$ recovers the original ordering: $\pi^{-1}(\hat{x}) = x$.

In pretraining, the model receives $\hat{x}$ and must predict $\pi^{-1}$ to reconstruct the original sequence. This prediction corresponds to one of $N!$ possible permutations. By completing this task, the model learns the natural spectral ordering and continuity characteristics found in real hyperspectral signatures.

\subsection{Model Architecture}

\begin{figure}[!htb]
\centering
\includegraphics[width=0.8\linewidth]{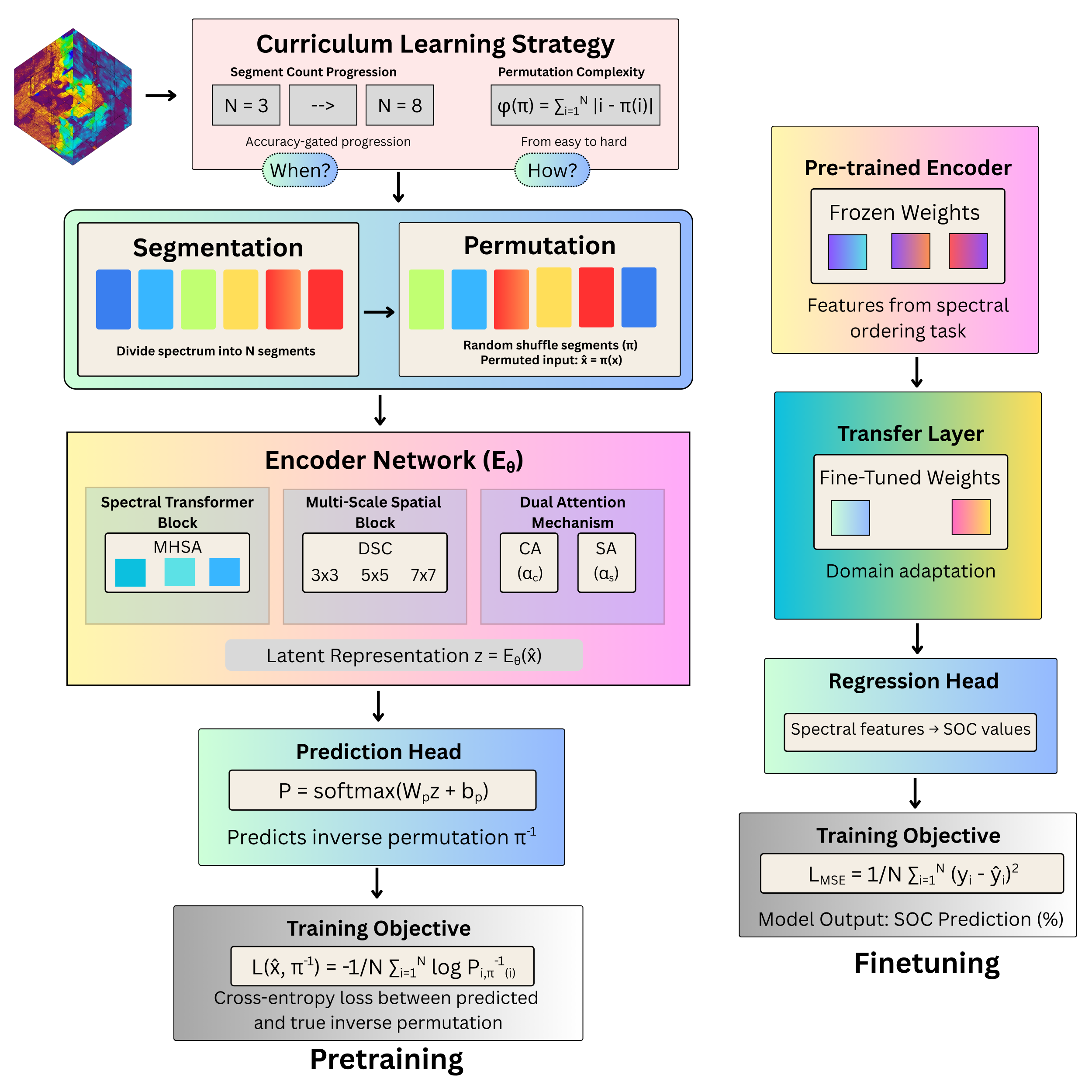}
\caption{Overview of the Spectral Band Permutation Prediction (SpecBPP) architecture. The pretraining phase (left) and the finetuning phase (right)}
\label{fig:specbpp}
\end{figure}

The SpecBPP architecture comprises an encoder network that extracts features from permuted spectral data and a prediction head that estimates the inverse permutation, as illustrated in Fig.~\ref{fig:specbpp}. 

\subsubsection{Encoder Network}

Our encoder $E_\theta$ processes the permuted spectral signature $\hat{x}$ to produce a latent representation $z = E_\theta(\hat{x}) \in \mathbb{R}^d$, where $d$ is the embedding dimension. As shown in Fig.~\ref{fig:specbpp}(left), the encoder consists of three principal components working in concert:

The \textbf{Spectral Transformer Block} employs a multi-head self-attention mechanism to capture dependencies across spectral bands \cite{Hong2022, Sun2022}. Formally, given input features $F \in \mathbb{R}^{H \times W \times B}$ (for spatial-spectral data with height $H$ and width $W$), the self-attention operation is defined as:
\begin{equation}
\text{Attention}(Q, K, V) = \text{softmax}\left(\frac{QK^T}{\sqrt{d_k}}\right)V
\end{equation}
where $Q$, $K$, and $V$ are query, key, and value projections. To incorporate domain knowledge about spectral relationships, we introduce a \textit{band weighting mechanism} $\alpha \in \mathbb{R}^B$ that adaptively emphasizes spectral regions known to correlate with soil properties:
\begin{equation}
F' = F \odot \alpha
\end{equation}
where $\odot$ represents element-wise multiplication broadcast along the spatial dimensions.

The \textbf{Multi-Scale Spatial Block} captures spatial context at multiple scales through efficient depthwise separable convolutions \cite{Howard2017, Chollet2017}. For each scale $s \in \{3,5,7\}$ (representing kernel sizes), we compute:
\begin{equation}
F_s = \text{DSConv}_s(F') = \text{PWConv}(\text{DWConv}_s(F'))
\end{equation}
where $\text{DWConv}_s$ is a depthwise convolution with kernel size $s$, and $\text{PWConv}$ is a pointwise (1×1) convolution. The multi-scale features are combined via:
\begin{equation}
F_{\text{MS}} = \text{Concat}(F_3, F_5, F_7)W + b
\end{equation}
where $W$ and $b$ are learnable parameters.

The \textbf{Dual Attention Mechanism} integrates spatial and channel attention to focus on informative regions \cite{Woo2018, Fu2019}. The channel attention $\alpha_c$ and spatial attention $\alpha_s$ are computed as:
\begin{align}
\alpha_c &= \sigma(W_2\delta(W_1\text{GAP}(F_{\text{MS}}))) \\
\alpha_s &= \sigma(f_{7\times7}(\text{MaxPool}(F_{\text{MS}}),\text{AvgPool}(F_{\text{MS}})))
\end{align}
where $\sigma$ is the sigmoid function, $\delta$ is ReLU, $\text{GAP}$ is global average pooling, and $f_{7\times7}$ is a 7×7 convolution. The final encoder output integrates these attention mechanisms:
\begin{equation}
z = \text{GAP}(F_{\text{MS}} \odot \alpha_c \odot \alpha_s).
\end{equation}

\subsubsection{Permutation Prediction Head}
The prediction head maps encoded representation $z$ to a distribution over possible inverse permutations. To handle the factorial growth of permutation space ($N!$), we adopt a factorized approach treating prediction as $N$ separate classification problems \cite{adams2011ranking, deshwal2022bayesian}.

The head outputs an $N \times N$ matrix $P$, where $P_{ij}$ represents the probability that segment at position $i$ originated from position $j$:
\begin{equation}
P = \text{softmax}(W_p z + b_p)
\end{equation}
where $W_p \in \mathbb{R}^{N \times N \times d}$, $b_p \in \mathbb{R}^{N \times N}$ are learnable parameters with softmax applied row-wise. The predicted inverse permutation is:
\begin{equation}
\hat{\pi}^{-1}(i) = \underset{j \in \{1,2,...,N\}}{\text{argmax}} \, P_{ij}.
\end{equation}
This approach efficiently decomposes the exponentially large permutation space into a tractable form.

\subsection{Training Objectives}

We train the model to minimize the cross-entropy loss between the predicted and true inverse permutation. Given a permuted spectrum $\hat{x} = \pi(x)$ and its inverse permutation $\pi^{-1}$, the loss is:
\begin{equation}
\mathcal{L}(\hat{x}, \pi^{-1}) = -\frac{1}{N} \sum_{i=1}^{N} \log P_{i, \pi^{-1}(i)}
\end{equation}
where $P_{i, \pi^{-1}(i)}$ is the predicted probability that the segment at position $i$ originated from position $\pi^{-1}(i)$.

This objective forces the model to learn the spectral continuity patterns necessary to unscramble the permuted segments. By correctly identifying the original positions, the model must develop representations that capture physically meaningful relationships between different regions of the electromagnetic spectrum. 

\subsection{Curriculum Learning Strategy}
To address the factorial explosion of permutation space ($N! $ possible permutations), we implement a curriculum learning strategy along two dimensions: segment count and permutation complexity.

For segment count progression, we define an adaptive six-phase curriculum from 3 to 8 segments:
\begin{equation}
\label{eq:curriculum}
N_t \;=\; 3 + \sum_{i=1}^{5} \mathbf{1}\bigl(\mathrm{val\_acc} \ge \alpha_i\bigr)
\end{equation}
where $\mathbf{1}(\cdot)$ is the indicator function and $\alpha_i$ are validation accuracy thresholds (99\%). This ensures mastery at each level before progressing to permutation spaces of increasing cardinality: 6, 24, 120, 720, 5\,040, and 40\,320.

Within each phase, we control permutation difficulty using function $\phi(\pi) = \sum_{i=1}^{N} |i - \pi(i)|$ that measures distance from identity \cite{sakaridis2019guided}. We bias sampling toward "easier" permutations initially, gradually increasing complexity:
\begin{equation}
p(\pi) \propto \exp\left(-\frac{\phi(\pi)}{T_s(t)}\right)
\end{equation}
where $T_s(t)$ increases with training progress.

Empirically, direct training with $N=8$ fails to converge, while our curriculum achieves 100\% validation accuracy for $N\leq7$ and 84.2\% for $N=8$, demonstrating its effectiveness for stable learning of spectral structure.

\section{Experiments}

We evaluate our SpecBPP approach on the task of Soil Organic Carbon (SOC) estimation using hyperspectral imagery. This section describes our experimental setup, presents comparative results against state-of-the-art methods, and provides analysis of the learned representations.

\subsection{Experimental Setup}
\paragraph{Datasets:} We use EnMAP satellite imagery (224 bands, 420-2450 nm, 30m resolution) \cite{storch2023enmap, chabrillat2024enmap}. For pretraining, we extract 196\,875 non-overlapping 64×64 patches from 1\,000 scenes. For fine-tuning, we use patches corresponding to 1\,540 soil samples with laboratory-measured SOC (0.5-23.8\%), split into training (70\%), validation (15\%), and test (15\%) sets with stratified sampling.

\paragraph{Baseline Methods:} We compare against: (1) traditional methods (PLSR, RF, SVR); (2) \textit{Supervised}: identical architecture trained only on labeled data; (3) \textit{MAE}: masked autoencoder for HSI \citep{he2022masked}; (4) \textit{I-JEPA}: joint-embedding architecture \citep{assran2023ijepa}; and (5) \textit{SimCLR}: contrastive learning \citep{chen2020simple}.

\paragraph{Evaluation Metrics:} We use coefficient of determination (R$^2$), root mean square error (RMSE), mean absolute error (MAE), and ratio of performance to deviation (RPD). RPD values above 3.0 indicate excellent performance for agricultural applications \cite{Chang2001}.

\paragraph{Implementation:} Models are implemented in PyTorch and trained on A100 GPUs with batch size 256. We train SpecBPP for 200 epochs (SGD, initial lr=$1\times10^{-3}$, cosine decay) and fine-tune for 150 epochs (lr=$5\times10^{-4}$) with early stopping on validation R$^2$.

\subsection{Results}
\paragraph{Pretraining Performance.} Table \ref{tab:permutation_accuracy} compares permutation prediction accuracy between direct training and our curriculum strategy across different segment counts. Our curriculum approach achieves 100\% accuracy on increasingly complex tasks (3→7 segments) by progressively advancing through segment counts upon mastery. For the most challenging case with 8 segments (40,320 possible permutations), the curriculum strategy still maintains strong performance with 84.2\% accuracy, significantly outperforming direct training which only achieves 4.9\% after the same number of epochs. This performance gap validates the effectiveness of our curriculum approach for complex permutation tasks.

\begin{table}[!htb]
\caption{Permutation prediction accuracy (\%) on validation data for different segment counts and training strategies.}
\label{tab:permutation_accuracy}
\centering
\scalebox{0.6}{
\resizebox{\textwidth}{!}{%
\begin{tabular}{ccccccc}
\toprule
\multirow{2}{*}{Training Strategy} & \multicolumn{6}{c}{Number of Segments ($N$)} \\
\cmidrule{2-7}
 & 3 & 4 & 5 & 6 & 7 & 8 \\
\midrule
Direct (Epoch 30) & 100.0 & 94.8 & 67.4 & 35.7 & 9.2 & 0.3 \\
Direct (Epoch 100) & 100.0 & 100.0 & 99.5 & 70.2 & 32.5 & 1.7 \\
Direct (Epoch 200) & 100.0 & 100.0 & 100.0 & 87.6 & 68.3 & 4.9 \\
\midrule
Curriculum (Epoch 30) & 100.0 & - & - & - & - & - \\
Curriculum (Epoch 50) & - & 100.0 & - & - & - & - \\
Curriculum (Epoch 90) & - & - & 100.0 & - & - & - \\
Curriculum (Epoch 140) & - & - & - & 100.0 & - & - \\
Curriculum (Epoch 170) & - & - & - & - & 100.0 & - \\
Curriculum (Epoch 200) & - & - & - & - & - & 84.2 \\
\bottomrule
\end{tabular}
}
}
\end{table}

\begin{figure}[!htb]
\centering
\includegraphics[width=0.7\textwidth]{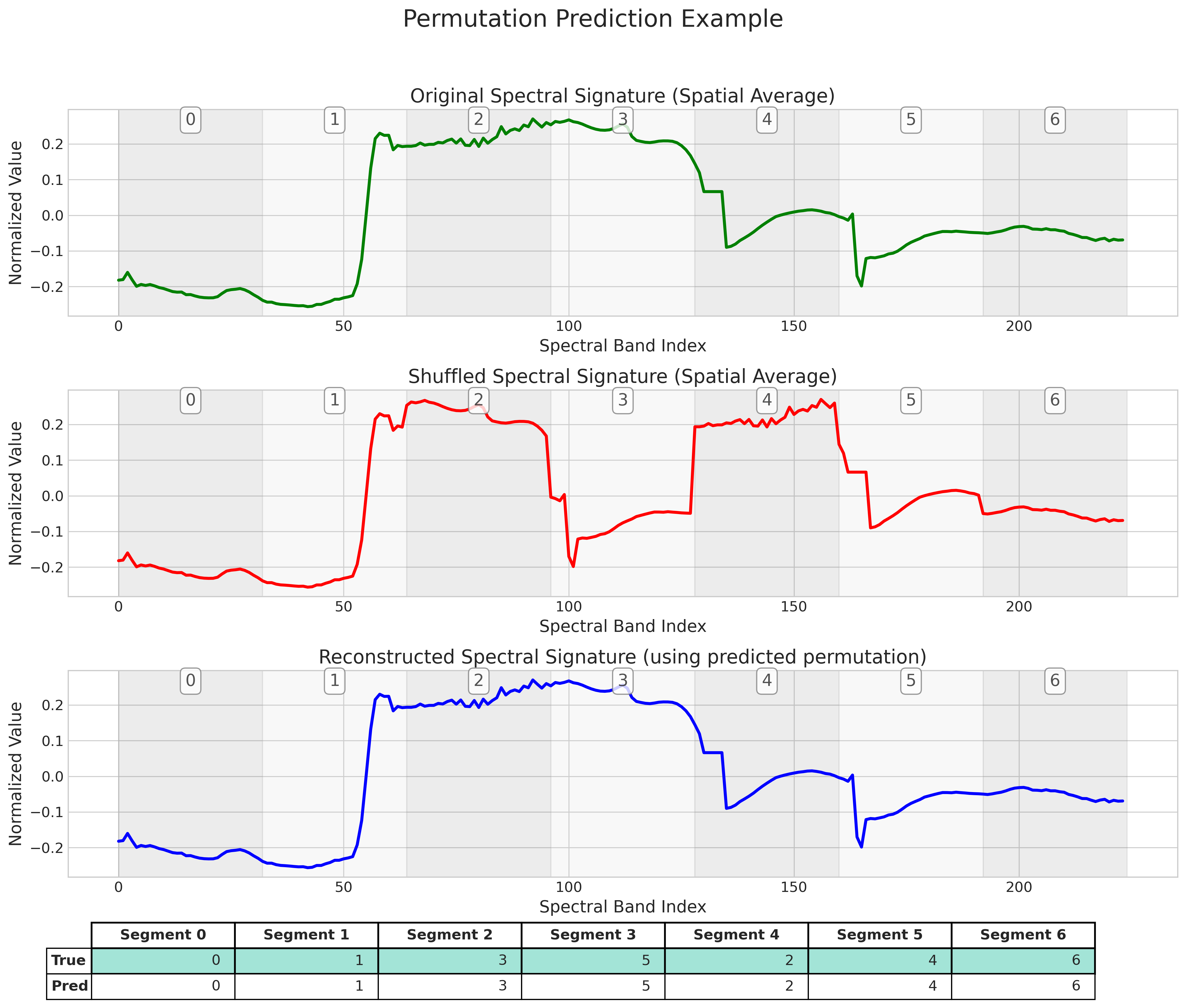}
\caption{Example of spectral band permutation prediction. Each panel shows: (top) the original spectral signature with 7 segments, (middle) the shuffled signature after permutation, and (bottom) the reconstructed signature after applying the predicted inverse permutation. The table shows the true vs. predicted segment positions, demonstrating perfect prediction accuracy.}
\label{fig:perm_example}
\end{figure}

Figure~\ref{fig:perm_example} demonstrates the spectral permutation prediction process. An original spectral signature with 7 segments (top) is randomly shuffled (middle), disrupting characteristic absorption features. The model correctly predicts the inverse permutation, reconstructing the original signature (bottom) with 100\% segment placement accuracy. Further visualizations of embeddings and full image band reconstructions are provided in the Supplementary Material.

\paragraph{SOC Estimation Performance.} Table \ref{tab:soc_performance} presents SOC estimation results after fine-tuning. 
SpecBPP performance improves consistently with segment count up to $N=7$ (R$^2$=0.9456, RMSE=1.1053\%, MAE=0.7394\%, RPD=4.1886), which significantly outperforms all baselines. Notably, performance decreases at $N=8$ (R$^2$=0.9033), suggesting an optimal segment count beyond which the factorial complexity becomes prohibitive.
The scatter plot provided in the supplementary material shows the quality of predictions is consistent across the entire range of SOC values, with detailed error analysis and performance visualizations available. 
Additional results and visualizations are provided in the Supplementary Material, including t-SNE visualization of the learned embeddings, detailed error analysis, and visual demonstration of spectral band permutation and reconstruction.

\begin{table} 
\caption{SOC estimation performance comparison. Bold values indicate best performance}
\label{tab:soc_performance}
\centering
\scalebox{0.6}{
\resizebox{\linewidth}{!}{%
\begin{tabular}{lccccc}
\toprule
Method & Pre-train & $R^2$ (↑) & RMSE (\%) (↓) & MAE (\%) (↓) & RPD (↑) \\
\midrule
PLSR & \xmark & 0.4823 & 3.8126 & 2.9843 & 1.3903 \\
RF & \xmark & 0.5010 & 2.7944 & 2.0871 & 1.4156 \\
SVR & \xmark & 0.4952 & 2.8103 & 2.1067 & 1.3794 \\
\midrule
Supervised & \xmark & 0.6231 & 2.3792 & 1.3654 & 2.2276 \\
MAE & \cmark & 0.8145 & 1.7218 & 1.0354 & 3.0783 \\
I-JEPA & \cmark & 0.8630 & 1.6543 & 0.9876 & 3.2037 \\
SimCLR & \cmark & 0.8073 & 1.7864 & 1.0892 & 2.9670 \\
\midrule
SpecBPP ($N=3$) & \cmark & 0.6028 & 2.5629 & 1.9203 & 1.6908 \\
SpecBPP ($N=4$) & \cmark & 0.7236 & 2.1347 & 1.4829 & 2.4651 \\
SpecBPP ($N=5$) & \cmark & 0.8345 & 1.6694 & 1.0147 & 3.1045 \\
SpecBPP ($N=6$) & \cmark & 0.8978 & 1.3421 & 0.8576 & 3.7329 \\
SpecBPP ($N=7$) & \cmark & \textbf{0.9456} & \textbf{1.1053} & \textbf{0.7394} & \textbf{4.1886} \\
SpecBPP ($N=8$) & \cmark & 0.9033 & 1.2987 & 0.8102 & 3.8452 \\
\bottomrule
\end{tabular}
}
}
\end{table}

\begin{table} [t] 
\caption{Ablation study of key components in SpecBPP ($N=7$).}
\label{tab:ablation}
\centering
\scalebox{0.6}{
\resizebox{\textwidth}{!}{%
\begin{tabular}{lccc}
\toprule
Model Configuration & $R^2$ (↑) & RMSE (\%) (↓) & RPD (↑) \\
\midrule
SpecBPP (Full) & \textbf{0.9456} & \textbf{1.1053} & \textbf{4.1886} \\
SpecBPP w/o CL & 0.9021 & 1.6632 & 3.1865 \\
SpecBPP w/o SW & 0.9318 & 1.3892 & 3.8151 \\
SpecBPP w/o DA & 0.9402 & 1.2147 & 3.8223 \\
\bottomrule
\end{tabular}
}
}
\end{table}

\paragraph{Ablation Studies.} Table \ref{tab:ablation} presents ablation results for key components of SpecBPP. Removing curriculum learning (CL) produces the largest performance drop (R$^2$ decreases by 0.0435), confirming its crucial role in managing permutation complexity. Spectral band weighting (SW) and dual attention (DA) also contribute positively but with less impact. 

\section{Conclusion}
We have presented Spectral Band Permutation Prediction (SpecBPP), a novel self-supervised framework that leverages the natural ordering of hyperspectral bands to learn globally coherent spectral representations. Our curriculum learning strategy helps SpecBPP overcome the factorial complexity of segment ordering and yields rich embeddings without any manual labels.

When fine-tuned on limited soil organic carbon (SOC) samples, our approach achieves state-of-the-art performance (R$^2$ = 0.9456, RMSE = 1.1053 \%, RPD = 4.19), substantially surpassing masked autoencoders, contrastive methods, and traditional regressors.

SpecBPP opens new avenues for self-supervised learning in remote sensing by explicitly encoding spectral continuity and long-range dependencies. 

\subsection{Limitations and Future Work}

Despite the promising results, SpecBPP has several limitations.  The factorial growth of permutation space ($N!$) imposes computational constraints, with accuracy decreasing from 100\% at $N=7$ to 84.2\% at $N=8$. Fixed segmentation strategies may inadvertently split meaningful absorption features across segment boundaries, limiting the spectral interpretation.  Our validation is currently limited to EnMAP imagery, with cross-sensor performance across various spectral characteristics still untested. While effective for estimating SOC, its application to other domains such as vegetation phenology, water quality assessment, and mineral mapping requires additional validation. Future research will look into hierarchical segmentation approaches, cross-sensor transfer learning, and integrated spatial-spectral pretext tasks to address these limitations and broaden SpecBPP's applicability across a wide range of hyperspectral remote sensing applications.

\newpage
\bibliographystyle{unsrtnat}
\bibliography{egbib}
\end{document}